\documentclass[aip,reprint,nofootinbib]{revtex4-1}

\usepackage[utf8]{inputenc}
\usepackage{amsmath}
\usepackage{amsfonts}
\usepackage{amssymb}
\usepackage{graphicx}
\usepackage{verbatim}
\usepackage{soul}
\usepackage{xcolor}

\begin{document}

\title[The electroluminescent properties based on bias polarity of the epitaxial graphene/aluminium SiC junction]{The electroluminescent properties based on bias polarity of the epitaxial graphene/aluminium SiC junction}

\author{M. Rejhon}
 \email{rejhonm@karlov.mff.cuni.cz.}
\author{J. Franc}
\author{V. D\v{e}di\v{c}}
\author{P. Hlídek}
\author{J. Kunc}

\affiliation{ 
Institute of Physics, Faculty of Mathematics and Physics, Charles University, Ke Karlovu 5, Prague, 121 16, Czech Republic}%

\date{\today}

\begin{abstract}

We investigated the electroluminescent properties of the epitaxial graphene/SiC junction. The temperature and current dependence of electroluminescence from the epitaxial graphene/SiC is measured within a temperature range of 50-300~K. The result of electroluminescence at 300~K is compared with the electroluminescent spectra from aluminium/SiC junction. The difference between the spectra is explained by the different band bending, which could lead to the tunable LED due to the semi-metal character of the graphene. We observed the electroluminescence at both bias polarities and we described the blue shift in the spectra by Franz-Keldysh effect.

\end{abstract}


\maketitle 


Silicon carbide is a material with an indirect broad bandgap from 2.3 to 3.3 eV depending on the SiC polytype. The unique properties of SiC, such as high saturation velocity of charge carriers or high thermal conductivity, make it a promising candidate for high-power and high-temperature applications. \cite{casady1996}. 

The first LED was made of SiC material in 1927 \cite{Zheludev2007}.  Jan et al. \cite{jan2009} investigated blue electroluminescence from an SiC tunneling diode. Strong electroluminescence in blue range was observed by Lebedev et al. \cite{lebedev1997}. Kamiyama studied the extremely high quantum efficiency of donor-acceptor-pair emission between nitrogen (donor) - boron (acceptor) and nitrogen - aluminium (acceptor) in SiC in a range of yellow and blue color \cite{kamiyama2006}, which makes SiC suitable for the production of white LED with a very high color rendering index. \cite{ou2014}  The color rendering index is a quantitative measure of the ability of a light source to reveal the colors of various objects faithfully in comparison with an ideal or natural light source. Optical emission between N and B has a wavelength at $590$~nm with full width at half maximum (FWHM) $\approx110$~nm and between N and Al has a wavelength at $460$~nm with FWHM $\approx80$~nm. Another prospective use of SiC is in the area of room temperature single photon emitters \cite{Fuchs2013}.  

Graphene is a two-dimensional crystal, consisting of a hexagonal lattice of carbon atoms. One of the methods of graphene growth is a thermal decomposition of SiC. The graphene layer is formed on the SiC surface. This growth method provides, besided chemical vapour deposition, the highest figure of merit in terms of graphene quality and scalability, the two requirements for applications in electronics and optoelectronics \cite{kunc2014,dong2014,berger2004}. There has been both intense theoretical \cite{inoue2012,kageshima2013} and experimental \cite{ohta2010,sun2011,borysiuk2012,ostler2013} research of graphene/SiC interface. However, detailed studies of electroluminescent properties of graphene/SiC interface are rare \cite{Anderson2012}.

We have studied electroluminescent (EL) properties of the graphene/SiC LED at different temperatures (50-300~K). We compared the results of an etalon sample with an Al/SiC interface and discussed differences of electrooptical properties. We verified in comparison with aluminium contacts, that the electroluminescence can be altered by different electron affinity of graphene. The altered EL emission is a fingreprint of surface-related electron-hole recombination. We show, employing graphene semimetalicity and low density of states, can be used for applications in tunable EL diodes. We present a theory based on the change of the bangap with the high electric inner field which explains the observed EL shift depending on the applied voltage.

We grew graphene on a Si-face of the conductive SiC 4H polytype (manufactured by II-VI Incorporation) by thermal decomposition of SiC at 1650$^{\circ}$C for 5 minutes in an argon atmosphere. SiC substrate was doped by nitrogen ($N_{D}\approx10^{16}$~cm$^{-3}$) and its dimensions were $3.30\times3.70\times0.35$~mm. The presence of graphene on the SiC surface was verified by Raman spectroscopy.

\begin{figure}[hbtp]
\centering
\includegraphics[width=6cm]{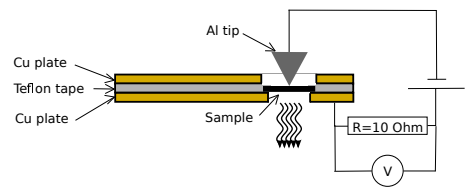}
\caption{Schematic cross sectional view of the sample.}
\label{vzorek}
\end{figure}

To measure EL the sample was inserted between a copper plate and Teflon tape. An aluminium tip was attached on the graphene layer (Figure \ref{vzorek}). The tip and Cu plate were connected to a current source Keithley 2400.

We have measured  current-voltage (I-V) characteristics and EL spectra in the temperature range $50-300$~K. The current source was Keithley 2400 which supplied current up to $\pm1$~A. Multimeter Keithley 2000 measured voltage ($U_{r}$) on the resistor $10\Omega$. The voltage on the sample is equal to $U_{s}=U-U_{r}$.
 
The spectrometer Princeton Instruments SpectraPro 2300i with grating 600 lines/mm and cooled CCD detector at a temperature of liquid nitrogen was used to measure the EL spectra. The EL was collected from the opposite side to which the Al tip was attached.

\begin{figure}[hbtp]
\centering
\includegraphics[width=8.5cm]{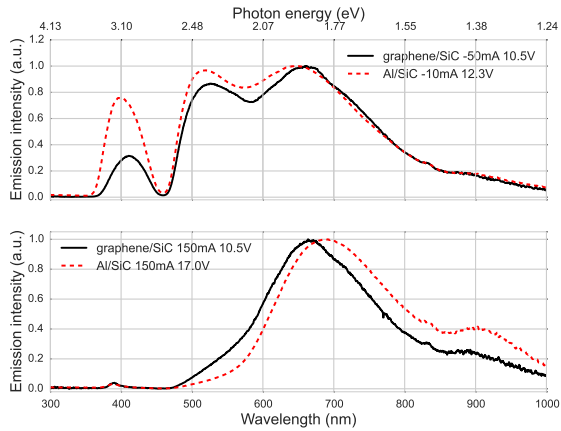}
\caption{Comparison of EL spectra of SiC with (black curve) and without graphene (red dashed curve). The top graph shows The EL in the negative polarity. The  EL spectra in the positive polarity is showed in bottom graph.}
\label{porovnani}
\end{figure}

We measured the EL spectrum from the sample with and without graphene contact. These spectra are shown in Figure \ref{porovnani}. It is evident that EL was observed in both polarities. The spectra of the sample with and without graphene look similar, but there is a difference between intensity and there is a small shift ($\Delta\lambda\approx10$~nm), which is caused by a different internal electric field. The internal electric field depends on band bending and on the work function of the metal. This means that if we change the work function of the metal, the EL spectra will change too. Fermi level in graphene is easily controlled by gate voltage because graphene is a semi-metal \cite{Yang1220527}. 

\begin{figure}[hbtp]
\centering
\includegraphics[width=8.5cm]{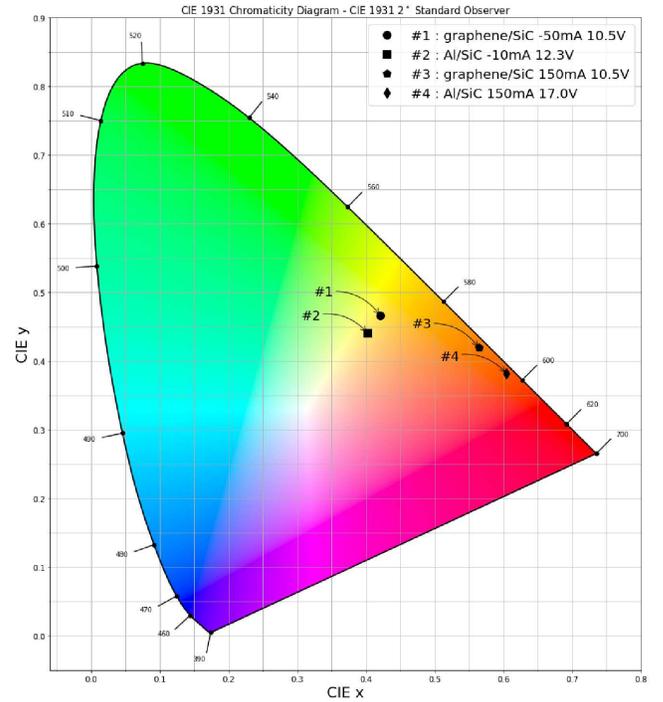}
\caption{Chromaticity diagram with color coordinates for sample emission at both polarities.}
\label{CIE}
\end{figure}

To illustrate the color in both polarities and both contacts the calculated color (CIE x,CIE y)-coordinates are shown on the Commision Internationl de l'Eclaire (CIE) 1931 chromaticity diagram on Figure \ref{CIE}.

\begin{figure}[hbtp]
\centering
\includegraphics[width=6cm]{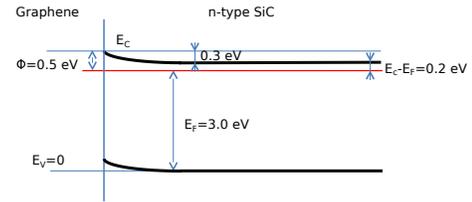}
\caption{Energy diagram of graphene/SiC structure without applied voltage. Idealized model of work function difference.}
\label{bending}
\end{figure}

The electron affinities of graphene and SiC are 4.5 and 4.0 eV, respectively \cite{Nagashio2010,Sze2007}. The 4H-SiC bandgap is 3.23~eV and the Fermi level is located below the conduction band. This means that the work function of SiC ($4.2$~eV = sum of electron affinity $4.0$~eV and $0.2$~eV energy difference between the bottom of the conduction band and the Fermi energy) is close to the work function of graphene. Therefore, a relatively small band bending upwards can be formed on the interface based on the simple model of work function difference (Figure \ref{bending}).  The band bending of aluminium/SiC interface is 0.1~eV due to the work function of aluminium, which is approximately 4.3~eV. 

\begin{figure}[hbtp]
\centering
\includegraphics[width=8.5cm]{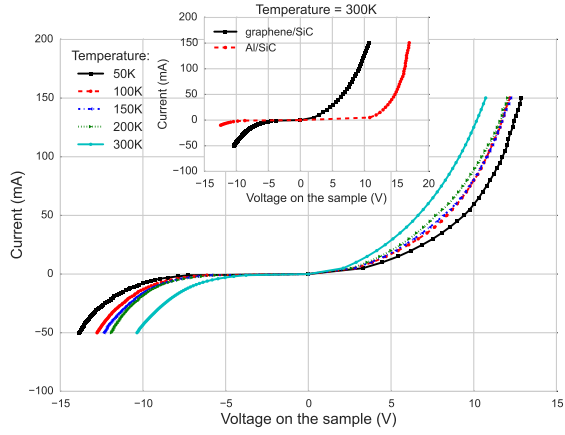}
\caption{The I-V characteristic of the measured graphene/SiC structure at different temperatures. The insert graph shows the difference between sample with aluminium contacts and with graphene contact.}
\label{IV}
\end{figure}

The temperature dependence of the sample I-V characteristic is shown in Figure \ref{IV}. The I-V curves were measured between $-50$~mA and $150$~mA with a step of $1$~mA for negative polarity and $5$~mA for positive polarity on Al tip.  The maximal voltage on the sample was about $13$~V for both current polarities. The current increases as the temperature increases due to the thermally excited electrons over the barrier of the graphene/SiC interface. The insert graph shows the difference between Al and graphene contacts. It is evident that graphene contact allows the carriers to move easily, in bulk, over the barrier.

\begin{figure}[hbtp]
\centering
\includegraphics[width=8.5cm]{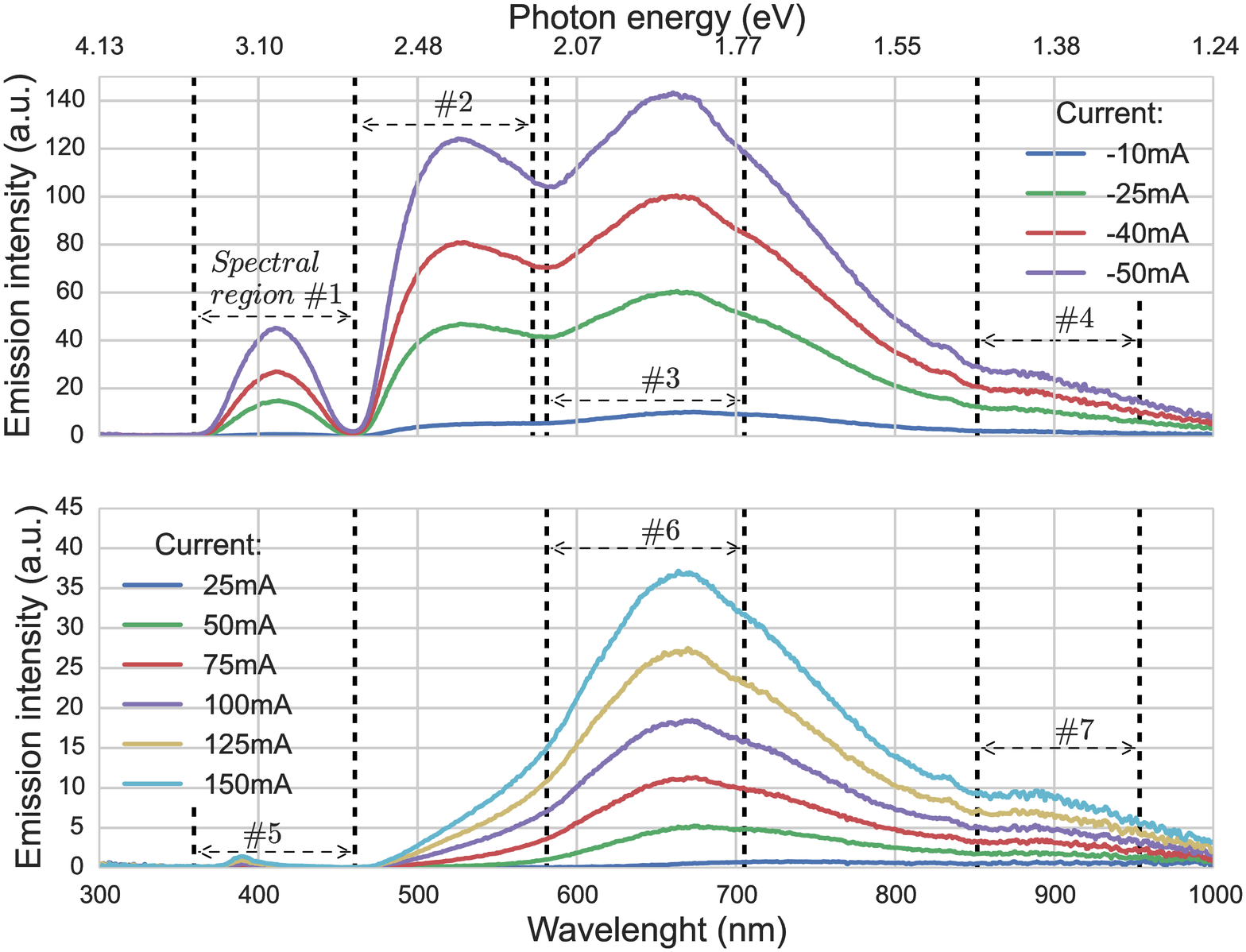}
\caption{Current dependence of graphene/SiC EL spectra at $300$~K. The spectra in negative polarity is shown in top graph. Bottom graph shows the spectra in positive polarity. The "band center of mass" is determined from the marked areas.}
\label{spectra}
\end{figure}

The current dependence of the EL spectrum at $300$~K is shown on Figure \ref{spectra}. The top graph shows the EL spectra in negative polarity (negative bias at the Al contact), which consists of four broad bands. $Spectral$ $region$ $\#1$ at $410$~nm ($3.02$~eV), $spectral$ $region$ $\#2$ at $520$~nm ($2.38$~eV), $spectral$ $region$ $\#3$ at $660$~nm ($1.88$~eV) and $spectral$ $region$ $\#4$ at $900$~nm ($1.38$~eV). The EL radiation was observed by naked eye. 

The EL spectra in positive polarity (plus on the Al contact) are shown in the bottom graph on Figure \ref{spectra}. There are three spectral regions, $spectral$ $region$ $\#5$ at $400$~nm ($3.10$~eV) with low intensity, broad strong $spectral$ $region$ $\#6$ at $660$~nm ($1.88$~eV) and $spectral$ $region$ $\#7$ at $900$~nm ($1.38$~eV).
The EL spectrum in positive polarity is approximately three times weaker than in negative polarity, but EL light still visible by naked eye.

\begin{figure}[hbtp]
\centering
\includegraphics[width=8.5cm]{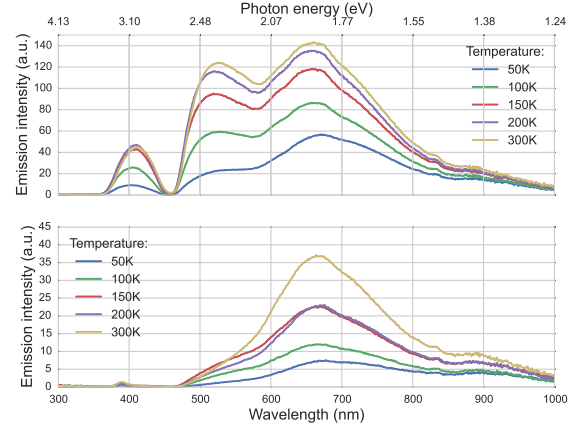}
\caption{Temperature dependence of graphene/SiC EL spectra. The spectra excited by $-50$~mA is shown in top graph. Bottom graph shows the spectra excited by current $150$~mA.}
\label{Tspectra}
\end{figure}

The temperature dependence of EL in forward (current set to 150 mA) and backward (current set to -50 mA) biased junction is plotted in Figure \ref{Tspectra}. The characteristic temperature dependence leads us to the conclusion that all the radiative transitions are caused by donor-acceptor pairs recombination because the EL intensity decreased with decreasing temperature due to the thermally activated electron-hole recombination between acceptor and donor levels.

The spectral regions at $3.02$~eV and $3.10$~eV are caused by the donor-acceptor transition between N and Al \cite{KALININA1997259,KUZNETSOV1995}. The spectral region at $2.38$~eV is ascribed to donor-acceptor-pair recombination between N donors and Al acceptors levels, too \cite{jan2009}.  The transitions between N donors and B acceptors is responsible for EL at $1.88$~eV \cite{kamiyama2006}. The last spectral region at $1.38$~eV could have its origin from a radiative transition between vanadium levels \cite{pavesi2004} or the transition between energy levels of a silicon vacancy \cite{Baranov2011}.

All radiative transitions take place in close proximity to the graphene/SiC interface where the tip is attached because there is the highest voltage gradient. When the negative polarity is applied, the electrons are injected from the graphene to the SiC and they can be trapped at nitrogen levels. The minority holes are drawn from the volume to the interface, where they are captured on the acceptor levels and then they are radiatively recombined with electrons from nitrogen levels.

In applied positive bias, the electrons are injected from the back contact and attracted from volume to the interface. However, the holes are blocked at the interface  and pass through the barrier in a small quantity from the graphene. They are immediately trapped on the boron levels, which have greater capture cross section than the other levels. Then the captured holes recombine with electrons  bound on nitrogen donors and cause the EL at $1.88$~eV ($660$~nm) with FWHM about $0.44$~eV ($160$~nm).

\begin{figure}[hbtp]
\centering
\includegraphics[width=8.5cm]{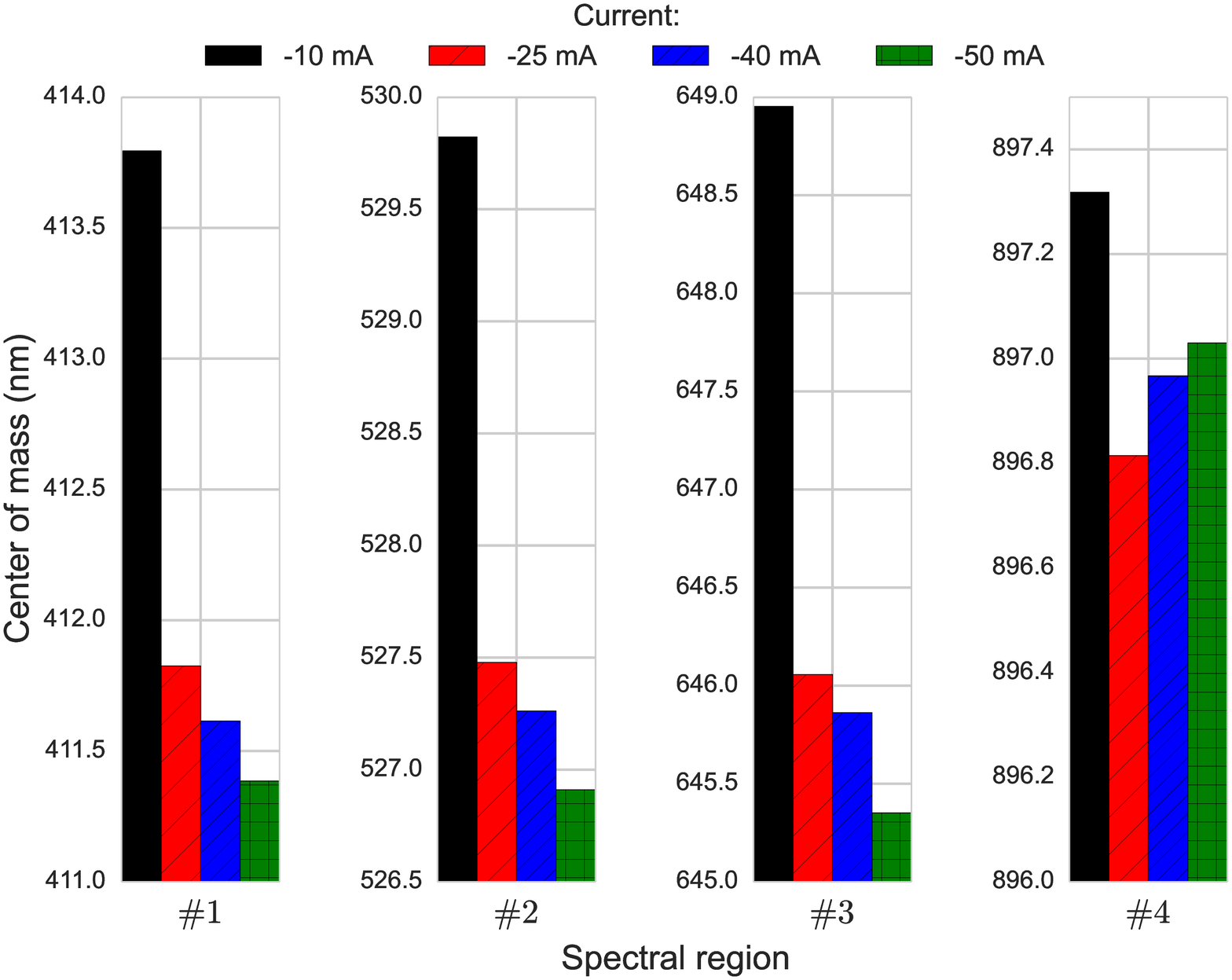}
\caption{Bar chart of the center of mass depending on the applied current at the sample.}
\label{bar}
\end{figure}

\begin{figure}[hbtp]
\centering
\includegraphics[width=8.5cm]{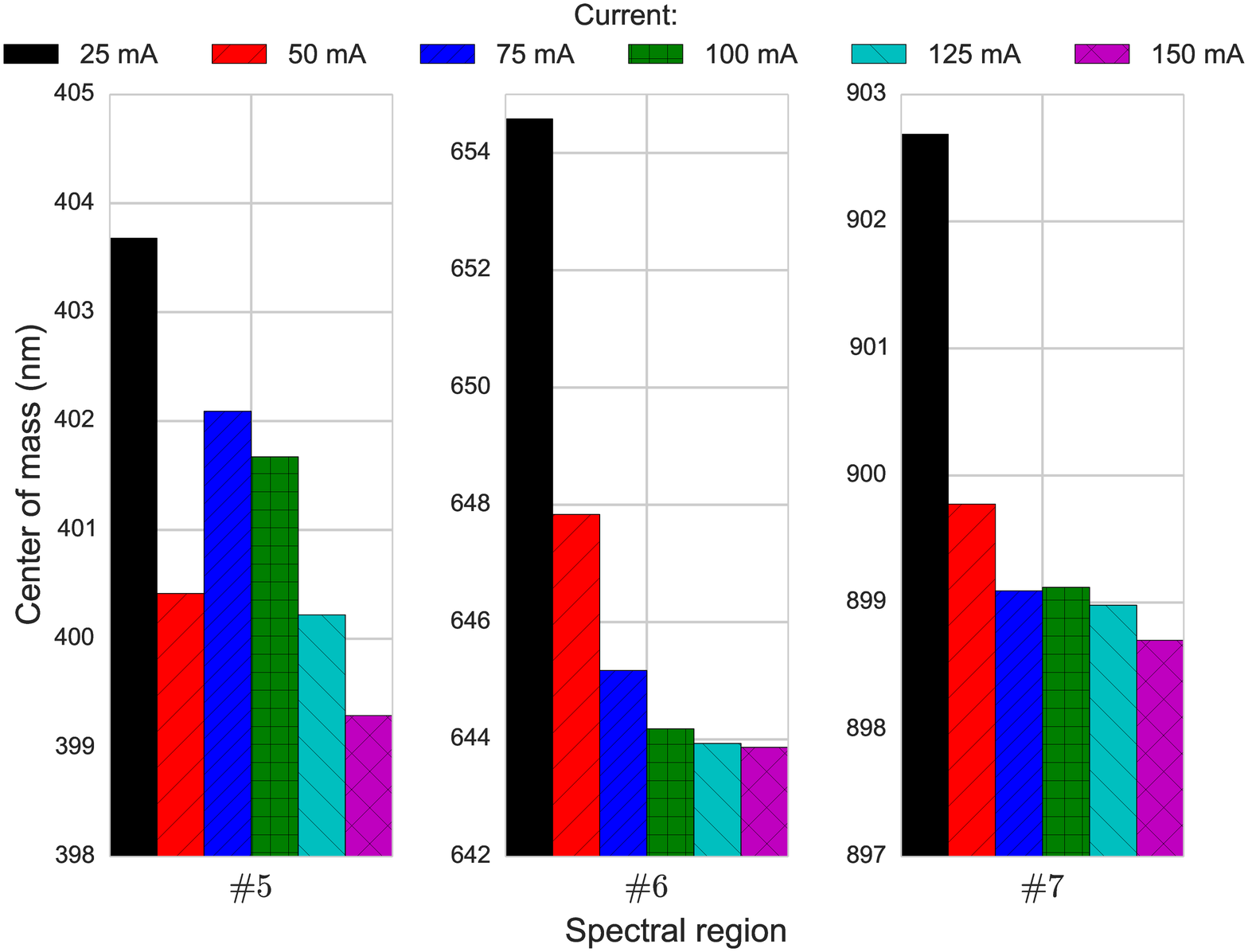}
\caption{Bar chart of the center of mass depending on the applied current at the sample.}
\label{bar2}
\end{figure}

\begin{figure}[hbtp]
\centering
\includegraphics[width=8.5cm]{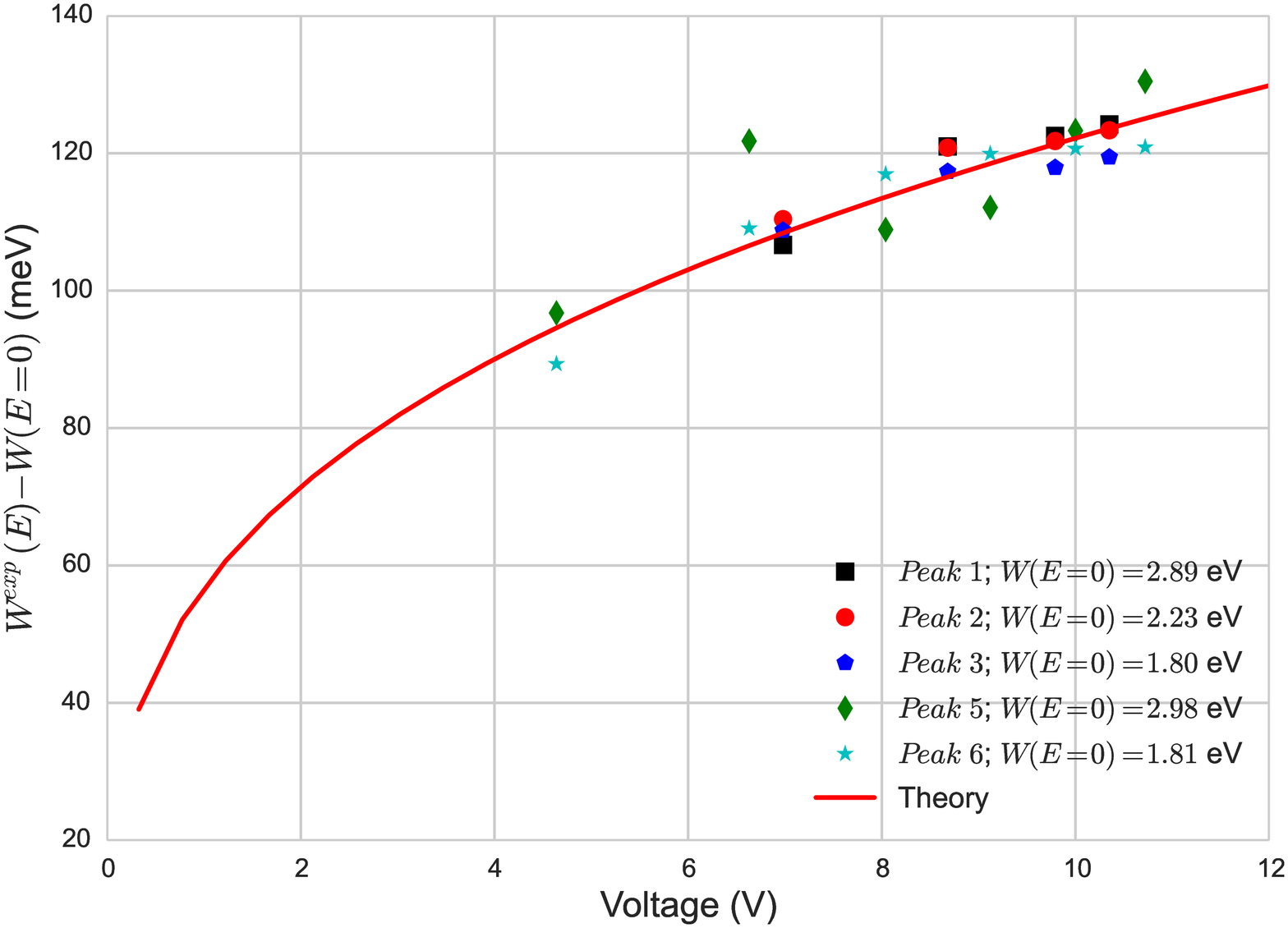}
\caption{Change of the bandgap depending on the applied voltage. The red line is the theory curve and the points are the experimental data minus the estimated spectral region centre of mass position without applied bias.}
\label{theory}
\end{figure}

The determined "band centers of mass" from marked areas in the graph on Figure \ref{spectra} is listed at Table \ref{Table1} for reverse polarity and at Table \ref{Table2} for positive polarity. The "band centers of mass" show a small blue shift (Figure \ref{bar} for reverse polarity and Figure \ref{bar2} for positive polarity). This shift is caused by the change of the bandgap in dependence on the electric inner field. The maximal electric inner field $E_{max}$ under interface is described by the formula \cite{Sze2007}

\begin{equation}
E_{max}=\sqrt{\frac{2eN_{D}V}{\varepsilon}},  
\end{equation}
where $V$ is the voltage, $N_{D}$ is a donor concentration, $e$ is the elementary charge and $\varepsilon$ is the absolute permittivity. The electric inner field is shifted in a range of $10^{7}-10^{8}$ V/m in our case of the applied voltage ($0-12$~V). The change of the bandgap $\Delta W^{exp}(E)$ depending on the electric internal field is described by Franz-Keldysh effect \cite{satzke1991} 
\begin{equation}
\Delta W(E)= W^{exp}(E)-W(0)=\frac{3}{2}\frac{(e\hbar  E_{max})^{2/3}}{(m^{\ast})^{1/3}},
\end{equation}
where $W(0)$ is the position of "band centres of mass" without applied bias, $\hbar$ is the reduced Planck constant and $m^{\ast}$ is the effective mass. The theoretical curve ($N_{D}\approx10^{16}$~cm$^{-3}$, $m^{\ast}=0.15m_{0}$, $\varepsilon\approx10\times\varepsilon_{0}$) and experimental data are plotted in graph in Figure \ref{theory} and there is a match between experimental data and theory. It means that the position of "spectral region center of mass" depends on the applied bias.

\begin{table}
\centering
\caption{Table of the determined center of mass for spectral band $\#1$ - $\#4$ depending on the applied current in the reverse polarity.}

\begin{tabular}{|c||c|c|c|c|}

\hline
  {\bf Current} & 
       \multicolumn{4}{c|}{\bf Spectral region} \\
\cline{2-5}
          (mA)    & $\#1$  & $\#2$ & $\#3$ & $\#4$ \\
\hline
\hline
  -10 & 413.79  & 529.82 & 648.95 & 897.32 \\
  -25 & 411.83  & 527.48 & 646.05 & 896.81 \\
  -40 & 411.61  & 527.26 & 645.86 & 896.97 \\
  -50 & 411.38  & 526.91 & 645.35 & 897.03 \\
\hline

\end{tabular}
\label{Table1}
\end{table}

\begin{table}
\centering
\caption{Table of the determined center of mass for spectral band $\#5$ - $\#7$ depending on the applied current in the positive polarity.}

\begin{tabular}{|c||c|c|c|c|}

\hline
  {\bf Current} & 
       \multicolumn{3}{c|}{\bf Spectral region} \\
\cline{2-4}
          (mA)    & $\#5$  & $\#6$ & $\#7$ \\
\hline
\hline
  25 & 403.68  & 654.58 & 902.69\\
  50 & 400.42  & 647.84 & 899.77\\
  75 & 402.09  & 645.17 & 899.09\\
  100 & 401.67  & 644.18 & 899.12\\
  125 & 400.22  & 643.93 & 898.97 \\
  150 & 399.30  & 643.86 & 898.70\\
\hline

\end{tabular}
\label{Table2}
\end{table}


In conclusion, the sample produces the EL in both current directions with a different emission spectral range. The observed EL spectra are formed by donor-acceptor transitions, which was confirmed by El intensity dependence on temperature. The blue shift of EL Spectral regions dependence on the bias was explained by the theory based on the change of bandgap due to the high electric inner field. We conclude that the control of the position of Fermi level in graphene by the voltage on the gate electrode can lead to a shift in the color perception of the emitted spectrum.

\section*{Acknowledgments}

The study was supported by the Charles University, project GA UK No.932216, the Grant Agency of Czech Republic under No. 16-15763Y and SVV-2017-260445. Raman spectroscopy has been measured as part of Project No. VaVpI CZ.1.05/4.1.00/16.0340.

\bibliographystyle{plain}

\end{document}